# Active causation and the origin of meaning

J. H. van Hateren

Johann Bernouilli Institute for Mathematics and Computer Science, University of Groningen, Groningen, The Netherlands

email: j.h.van.hateren@rug.nl

**Abstract** Purpose and meaning are necessary concepts for understanding mind and culture, but appear to be absent from the physical world and are not part of the explanatory framework of the natural sciences. Understanding how meaning (in the broad sense of the term) could arise from a physical world has proven to be a tough problem. The basic scheme of Darwinian evolution produces adaptations that only represent apparent ("as if") goals and meaning. Here I use evolutionary models to show that a slight, evolvable extension of the basic scheme is sufficient to produce genuine goals. The extension, targeted modulation of mutation rate, is known to be generally present in biological cells, and gives rise to two phenomena that are absent from the non-living world: intrinsic meaning and the ability to initiate goal-directed chains of causation (active causation). The extended scheme accomplishes this by utilizing randomness modulated by a feedback loop that is itself regulated by evolutionary pressure. The mechanism can be extended to behavioural variability as well, and thus shows how freedom of behaviour is possible. A further extension to communication suggests that the active exchange of intrinsic meaning between organisms may be the origin of consciousness, which in combination with active causation can provide a physical basis for the phenomenon of free will.

**Keywords** Emergence · Evolution of meaning · Agency · Consciousness · Free will

## 1 Introduction

Science maintains two different views on the world. The natural sciences see a physical world composed of energy and matter, whereas the fields studying mind and culture see a human world composed of purposeful behaviour and meaning (in the broad sense of the term, see Sect. 5). While the physical world is thought to be governed by mechanistic processes without purpose, the human world can only be understood when meaning and the active pursuit of goals are acknowledged as real phenomena. Darwin's theory of evolution showed how complex life-forms can evolve from simpler ones (Darwin 1859), and thus promised to connect the physical and human worlds. However, the origin of purpose and meaning is still not understood. The fundamental problem is that Darwinian evolution itself is a physical process, devoid of meaning. It is not clear if and how a physical process can generate phenomena that appear to be non-physical.

Here I present a toy model that generates meaning and goal-directedness as adaptations evolvable by Darwinian evolution. The model is intended as a proof of concept, not as a specific model of the processes involved. It contains the bare minimum required for Darwinian evolution, highlighting its causal structure. Lacking molecular details, the model should be equally applicable to chemically different life-forms that may have evolved on other planets than Earth (Bedau and Cleland 2010; van Hateren 2013).

Although the model only has direct empirical support from known molecular processes that regulate the mutation rate of biological cells, its basic mechanism (an evolutionary controlled feedback process modulating randomness) can be extended to shorter timescales than those of evolution. In particular, behavioural modifications that take place within the lifespan of an organism can be controlled by a similar mechanism, which thus leads to goal-directed behaviour with some degree of freedom, i.e., a form of agency. Extending the mechanism to communication between organisms suggests a plausible link with consciousness and free will.

The article is organized as follows, in order to enhance readability. The main ideas are presented primarily through text and figures in Sect. 2 ("Results"), Sect. 3 ("Discussion"), and in Figs. 1-3. Equations, and figures that illustrate or extend the main ideas (Figs. 4-10) are in Sect. 4 ("Methods"). Terminology is discussed in some depth in Sect. 5 ("Appendix"). In places in Sect. 2 and 3 where is referred to equations and figures in Sect. 4, this is not essential for following the line of argument, and the reader may prefer to skip these forward references on first reading.

## 2 Results

In this section the origin of meaning and active causation is first explained for hereditary change on an evolutionary timescale, with as unit the organismal lineage (here defined as a particular line of descent of organisms). Subsequently, these ideas are extended to behavioural change on the timescale of an organism's lifetime, with as unit the single organism. In the latter case, active causation takes the form of agency, the property of organisms that their behaviour is neither fully determined nor random, but changes in a way that is at the same time unpredictable and under control of the organism itself, thus providing the organism with a limited form of behavioural freedom (Heisenberg 2009; Brembs 2011).

### 2.1 Active causation of heredity

The basic Darwinian model of evolution (Fig. 1a) assumes a population of organisms, with $n(h,t)$ the number of organisms of hereditary type $h$ at time $t$. Each organism has a typical lifetime $\tau_H$ and produces, on average per lifetime, $f$ offspring, with $f(h,t)$ defined here as the organism's fitness. Note that fitness thus defined is a predictive, probabilistic variable, which varies during the lifetime of a particular organism, for example when circumstances deteriorate or improve. It can best be seen as an instantaneous reproductive rate, normalized by lifetime. The number of organisms of type $h$ is multiplied by $f$ and thus gradually changes. Without mutation, the number of type $h$ would rise exponentially as long as $h$ has $f$>1, whereas $f$<1 leads to decline and eventual extinction. The positive feedback loop (*I*) is the main engine of life because of its potential for exponential growth, but it is inherently unstable. A second, negative feedback loop (*II*) provides stabilization: because of limited resources, the fitness of all types declines when the total number of organisms $N$ approaches the environment's carrying capacity, for example when food or space becomes scarce. The fitness of some types $h$ consequently drops below one and the numbers of those types decrease, with steady or increasing numbers of the other types (natural selection). The result is a change in distribution of types $h$.

The model has two external inputs, an environmental variable $E$ and a mutation rate $\lambda_H$. $E(t)$ varies randomly over a wide range of temporal scales (Sect. 4, Fig. 4), both slower and faster than $\tau_H$. The fitness of type $h$ declines with the distance between $h$ and $E$ (Sect. 4, Eq. (2)), and $E(t)$ therefore drives changes in $n(h,t)$. Types $h$ can change into other types by mutation at a rate $\lambda_H$, spreading a particular type $h$ gradually to neighbouring types.

The organismal lineage has only a passive role in basic Darwinian evolution. It is driven by mechanistic feedback loops (*I* and *II*) and by external influences beyond its control ($E$ and stochastic processes producing mutations). Changes in fitness, heredity, and type distribution are produced, then, by forms of causation that are passive from the point of view of the organismal lineage. Moreover, the evolutionary process has no foresight or goal. When it produces organisms with adaptations matched to their current environment, it is because such adaptations happened to promote fitness in previous environments and $E(t)$ contains slow components remaining approximately the same over many $\tau_H$. Although adaptations may be perceived, post hoc, as goal-directed ones produced by a goal-directed process, either goal-directness is only apparent ("as if"). It is sometimes assumed that basic Darwinian evolution implies that high fitness and survival are genuine intrinsic goals of organisms, or even that self-maintenance as such is an intrinsic goal of life. However, this tacitly assumes that existing is better than not existing, a value judgment inapplicable to regular physical systems (Davies 2009), as they lack value to themselves.

The above analysis changes dramatically, however, when the basic Darwinian scheme is slightly extended. This extension provides the organisms of a lineage with some control over their own

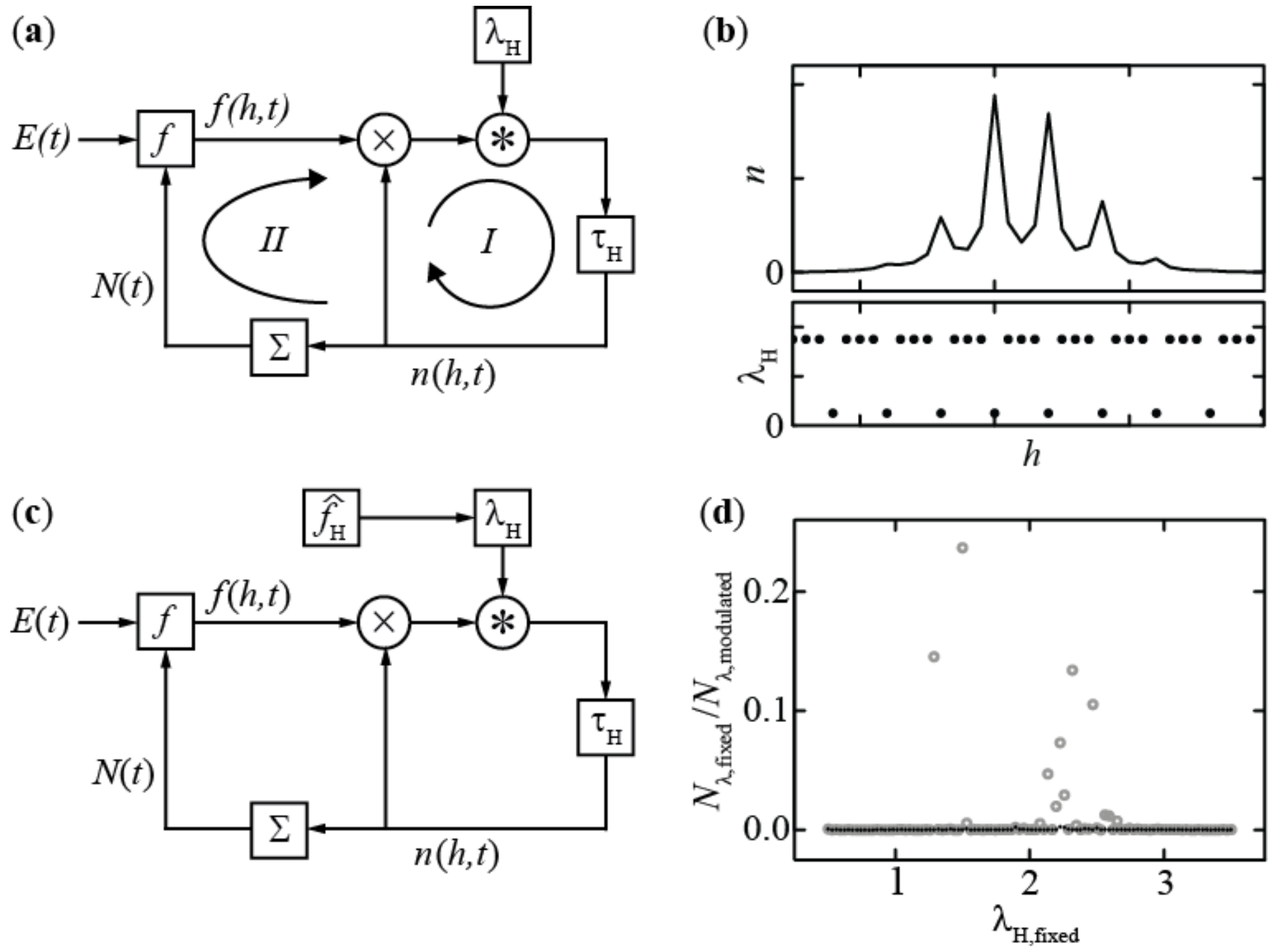

**Fig. 1** Population models of basic and extended Darwinian evolution. (**a**) Basic model. Fitness *f* modifies the number of organisms *n* multiplicatively (symbol ×), a mutation rate characterized by $\lambda_H$ modifies the distribution of types (convolution symbol ∗), and $\tau_H$ is the typical lifetime of an organism. See text for further explanation and Sect. 4.1 for details. (**b**) Modulating $\lambda_H$ as a comb function of *h* (lower panel) produces a comb-like population density (upper panel). The goal-function is such that one in four values of *h* has $\lambda_H$=0.5, and the others $\lambda_H$=3.5. The figure shows the result after 1000 time steps (10$\tau_H$), typical of the result for any time after an initial spin-up time of a few $\tau_H$. Labels at the tick marks (*h*: 10, 20, 30; $\lambda_H$=0, 2, 4; *n*: 0, 500, 1000) are partly suppressed for the sake of clarity. (**c**) Extended model. An estimate of its own fitness made by the organism, $\hat{f}_H$, is used to modulate $\lambda_H$. See text for explanation and consequences, and Sect. 4.2 for details. (**d**) Ratios of population numbers *N* of 100 pairs of competing populations, one always following the extended model and the other the basic model over a range of $\lambda_{H,fixed}$. Results after 300$\tau_H$ (open grey symbols) and 600$\tau_H$ (black dots). The results show that the extended model is evolvable

mutation rate. Such controls exist as a general feature of cells (Galhardo et al. 2007), observable as increased mutation rates in cells under stress, for example because of a shortage of nutrients or the presence of toxins. Although a higher mutation rate produces many nonviable offspring, it also increases the probability of a mutation that lets its carrier survive the stress, multiply, and thereby increase the chances that at least one of its lines of descent will survive (Ram and Hadany 2012). Note that this internal form of controlling mutation rate during the organism's lifetime is different from mutator alleles being selected (Taddei et al. 1997), which is externally driven (see further Sect. 3).

The consequences of a mutation rate under control of the organisms in a lineage can be understood from the following, hypothetical example. Suppose that the mutation rate is controlled by a function *g*, so it is given by $\lambda_H(g)$. In principle, *g* can take any form. Suppose it depends on heredity *h*, with *h* not only determining fitness, but also another property unrelated to fitness. For example, at regularly spaced positions along the *h*-axis organisms are slightly larger than for other *h*. If the former values of *h* produce a smaller $\lambda_H$ than the latter (Fig. 1b, lower panel), the population accumulates at the former ones because mutation away from those values is slower than from the other ones (Fig. 1b, upper panel). Even though the size variations produce no fitness change and the mutations are undirected (not changing *h* in a specific direction), there is a clear, directed effect on the distribution of types *h*. The function *g*, which is intrinsic to the organisms of a lineage, thus determines the probability of obtaining specific values of *h*. It can therefore be viewed as an implicit goal of the organism and its lineage, because $g$ influences the future heredity of the organismal lineage in a directed way.

The above example is unlikely to evolve because it does not increase fitness. Even if it would evolve by chance, it would be unsustainable on an evolutionary time scale, because it is at a disadvantage relative to lineages that do not spend resources on a property that is irrelevant to fitness. An evolvable and sustainable goal-function must improve fitness, and in its purest form must make $\lambda_H$ a function of an estimate $\hat{f}_H$ of the fitness of the organism, thus $\lambda_H(\hat{f}_H)$ (Fig. 1c). This is indeed an evolvable strategy: taking a mutation rate $\lambda_H$ that decreases as a function of estimated fitness $\hat{f}_H$ (high fitness implies low mutation rate, and *vice versa*, Sect. 4, Eq. (6)) produces a population of organisms that outperforms populations with $\lambda_H$ fixed (Fig. 1d). For each value of $\lambda_{H,fixed}$ in Fig. 1d, a simulation was performed with two competing populations, one population following the extended model (with modulated $\lambda_H$) and the other the basic model (with fixed $\lambda_H$). The ratio of the total number of organisms in each population fluctuates, but gradually the organisms of the extended model drive those of the basic model to extinction (thus the ratio $N_{\lambda,fixed}/ N_{\lambda,modulated}$ goes to zero, as illustrated by the ratio after 300 and 600 lifetimes, circles and dots, respectively; results are for 100 different simulations for a range of values of $\lambda_{H,fixed}$). The function $\hat{f}_H$ thus produces a goal-directedness in the lineage and its organisms that is itself evolvable by the basic Darwinian scheme. The goal evolves in the extended Darwinian scheme by piggybacking on the basic scheme, and aligning itself with it.

The true fitness $f$ is unlikely to be known exactly to a real-world organism (i.e., more realistic than the toy organisms modelled here), because for that it would need to make an accurate simulation of itself and its interactions with other evolving organisms and the environment. Although organisms indeed appear to spend a significant part of their physiological and neural resources on inferring the states of themselves and their environment, for example their internal nutritional state or dangerous external factors, producing the estimate $\hat{f}_H$ ultimately requires evolved heuristics (short-cuts and rules of thumb). Nevertheless, $\hat{f}_H$ is under selection pressure to provide an estimate as close to $f$ as possible, given the physiological constraints of the organism. The model calculations presented here assume that the toy organisms can make an estimate $\hat{f}_H$ that is arbitrarily close to $f$ (by knowing or observing their own type $h$, by having evolved a sensor that measures $E(t)$, and by approximating the mapping of Eq. (2) by an evolved physiological or neural model), but this is not a crucial assumption. Even if $\hat{f}_H$ is not optimal and deviates somewhat from $f$, similar results as in Fig. 1d are obtained (Sect. 4.2; see further Sect. 3 for discussion). Note that $f$ (and by implication $\hat{f}_H$) is not a constant but a function of time, changing continually because it depends on the momentary environment $E(t)$ and the variable population size $N(t)$ (Sect. 4, Eq. (2)). Both $f$ and $\hat{f}_H$ are predictive variables, related to an expected reproductive rate of each organism, with actually realized reproduction varying stochastically.

The extended Darwinian scheme not only produces a goal-directedness in the organisms of a lineage, but also changes the nature of the causality of the process. Causation in nature can be quite complex, but it is useful to distinguish two basic variants, deterministic and stochastic causation. In deterministic causation, systems evolve in a predictable way, with one part of the system influenced by and influencing other parts through fixed relationships. Chains of causation are mere continuations of existing chains, no new chains are started. In stochastic causation, new chains of causation are initiated by definition, because the stochastic variables have no specific (identifiable) cause. Physical sources of stochastic causation are thermal and quantum noise (see also Sect. 4), and thermal noise is indeed the primary source of stochasticity for mutations. The mutations as produced by $\lambda_H$ in the basic Darwinian scheme of Fig. 1a are the result of stochastic causation. In contrast, the modulated mutation rate $\lambda_H(\hat{f}_H)$ of the extended Darwinian scheme (Fig. 1c) is rather special, producing what might be called modulated stochastic causation. Effectively, it is equivalent to a non-negative deterministic variable ( $\hat{f}_H$ ) modulating the expected variance of the resulting mutations (in the toy model here along a single dimension, in reality in some high-dimensional space). Only the variance is modulated, not the mean, because modulating the mean would imply a directionality for which the organism would have no justification (the organism cannot foresee the effects of mutations on fitness).

The modulated stochastic causation produced by $\lambda_H(\hat{f}_H)$ mixes deterministic causation (through $\hat{f}_H$ driving variance) and stochastic causation (through random mutation). This would be rather uninteresting, were it not that it occurs in a feedback loop operating at the scale of an organismal lineage. The random mutations change an organism's descendant such that fitness is (usually) changed as well, such a fitness change is then reflected in the descendant's estimate of its own fitness, this self-estimated fitness subsequently influences the expected variance of mutations, the resulting mutations subsequently change the fitness of the next descendant in a lineage, and so on and so forth. The result is that the deterministic and stochastic causation become thoroughly entangled: fitness at a particular point in time is dependent on the complete fitness history of the lineage (and thereby on the history of the environment $E$) and on the complete history of the realizations of mutational stochasticity. Because mutational stochasticity depends on fitness, it therefore also depends, at a particular point in time, on the complete history of fitness, on $E(t)$, and on previous stochastic realizations. This is in stark contrast to the mutational stochasticity in the basic Darwinian scheme of Fig. 1a, which is not modified by fitness (i.e., not by fitness $f$ or $\hat{f}_H$ as the predictive, probabilistic variables defined here, although it may depend on fitness outcomes, actual realizations of fitness, for example when mutator alleles are promoted, in hindsight, by natural selection). The common assumption made in modelling that noise is independent from realization to realization (or has a limited correlation time as in coloured noise), and that it is independent of the signal, is thus strongly invalidated by this scheme. The effect is that fitness and mutations become strongly intertwined, and the causation should therefore be seen as a unique form, midway between deterministic and stochastic. The regular forms of deterministic and stochastic causation one finds in nature can be called “passive”, because things just happen as they do, either in a determined or spontaneous way. In contrast, the entangled form of causation described here will be called “active causation”, because of the crucial role the organisms of a lineage play in steering stochasticity and therefore in steering the resulting hereditary trajectories, as shown below. Hereditary trajectories describe the path that subsequent organisms in a lineage take through hereditary space. In Sect. 5 the use of the terms “active” and “passive” is discussed further.

In summary, the basic idea that underlies active causation is that mutation rates (resulting in random fluctuations) are actively determined by an estimated fitness function such that mutation rates are greater in areas of low expected fitness. Heuristically, this causes the dispersion of phenotypes from regimes with low fitness thereby ensuring that these regimes are avoided (like evacuation by local diffusion, as illustrated in Fig. 1b). This means that active causation can prescribe regions of fitness space that are more likely to be occupied – regions that have a low dispersion (a low mutation rate) because of their high estimated fitness.

The difference between causation in the basic and extended Darwinian schemes can be illustrated by simulating the evolution of lineages (Fig. 2a), and comparing how lineages would evolve under the two schemes. When a specific hereditary change $\Delta h$ occurs, it could be the result of either the extended Darwinian scheme ($\lambda_{H,modulated}$) or the basic one ($\lambda_{H,fixed}$). The likelihood ratio is the ratio of the probability of $\Delta h$ given modulated $\lambda_H$, $P(\Delta h|\ \lambda_{H,modulated})$, and the probability of $\Delta h$ given fixed $\lambda_H$, $P(\Delta h|\ \lambda_{H,fixed})$. This likelihood ratio is expected to be larger than one, on average, if $\Delta h$ was actually produced by the extended rather than the basic scheme, but for individual steps in $\Delta h$ the ratio is too variable to show the difference. However, for an organismal lineage the sequence of individual steps produces a total likelihood ratio that is the product of the likelihood ratios of individual steps. The likelihood ratio of hereditary trajectories ($^{10}$log shown in Fig. 2b) typically rises to high values, more quickly so at times when fitness is low. This means that $h$-trajectories produced by the extended scheme could not have been produced by organisms that keep $\lambda_H$ fixed. Such an $h$-trajectory is therefore partly, but observably, caused by how the organisms of the lineage estimate $\hat{f}_H$, and not any more only the result of passive causes as in the basic Darwinian scheme. Calling this form of causation active emphasizes this difference. Each step is random, and thus provides an element of unpredictability, but the trajectory as a whole is not random. It is the result of permanent changes in heredity continually guided by a feedback loop acting through $h$ and $\hat{f}_H$, and is thus dependent on previous (unpredictable) changes in *h* as well as on the goal function $\hat{f}_H$ . A trajectory is therefore neither completely predictable, because of the stochastic nature of mutation, nor completely unpredictable, because it is partly determined by each organism in the lineage through its estimate $\hat{f}_H$ of its own fitness. The computation of the likelihood ratio as performed here is not an essential part of the model, but simply serves to show that the influence of the goal function $\hat{f}_H$ on *h*-trajectories is not instantaneous, but gradually increases, up to a point where it is fully obvious.

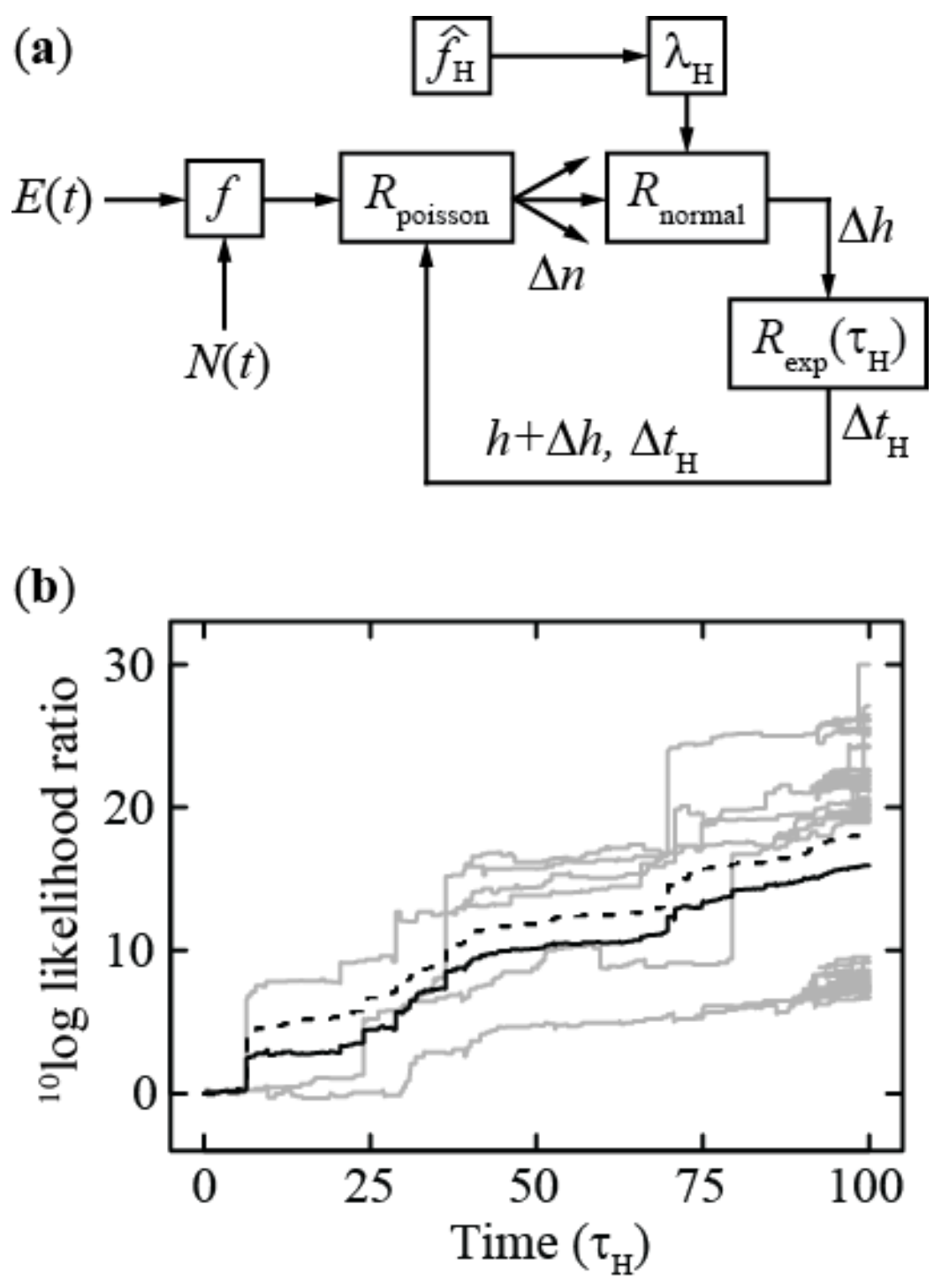

**Fig. 2** Organismal lineages are partly determined by $\hat{f}_H$ . (**a**) Equivalent of Fig. 1c for computing *h*-trajectories, with an *h*-trajectory the trajectory through hereditary space of a lineage, a line of descending organisms. *R* are random generators, $\Delta n$ is the realized number of offspring of an organism, $\Delta h$ the change in heredity (relative to the parent) of a particular descendant, and $\Delta t_H$ the lifetime of that organism. When $\Delta n$=0 the trajectory dies, when $\Delta n$ >1 it splits. For details see Sect. 4.3. (**b**) Likelihood ratios of *h*-trajectories $\Pi_{\Delta h}P(\Delta h|\ \lambda_{H,modulated})/\ P(\Delta h|\ \lambda_{H,fixed})$ typically rise to high values. Grey lines: 167 trajectories at end point, black line: mean, broken line: mean of theoretically expected likelihood ratios (Eq. (7)). The increasing likelihood ratios show that *h*-trajectories gradually become nearly completely determined by the goal function $\hat{f}_H$ , even if this only involves random hereditary changes. See text for further explanation and Sect. 4.4 for details

Because the organismal lineage has an active role in determining an *h*-trajectory, the intrinsic goal-directedness produced by $\hat{f}_H$ should be seen as a genuine goal. By internalizing the true fitness *f* as an estimate $\hat{f}_H$ , and utilizing that estimate in a feedback loop as explained above, the “as if” goal of basic Darwinian evolution is transformed into a genuine goal. Note, however, that this only applies to the intrinsic goal-directedness of the organisms in a lineage, the other parts of the evolutionary process remain invariably without goal. The above analysis shows that letting $\lambda_H$ depend on $\hat{f}_H$ has two consequences. First, it makes high fitness a genuine goal of the organisms in a lineage, and second, it gives them, through their lineage, an active role in its destiny. Actively pursuing a goal implies that reaching that goal is important to the organisms. The goal is embodied in the detailed form of $\hat{f}_H$ and is denoted here by the term “intrinsic meaning”. Meaning is used here not in the narrow sense (as in the meaning of a word) but in the broad sense (as in the meaning of an action or the meaning of life), implying significance, value, import, and purpose (see Sect. 5 for a discussion of this terminology).

## 2.2 Active causation of behaviour and dialogue

Hereditary variations only allow adaptations to variations of *E*(*t*) much slower than $\tau_H$. An organism can adapt to faster changes of *E*(*t*) as well by changing during its lifetime. Such changes are called here “behaviour”, used broadly, including changes that occur within cells and in immobile organisms. Active causation and intrinsic meaning are produced by analogy with the extended Darwinian scheme (Fig. 3a). Different behaviours *b* have a typical lifetime $\tau_B$ ($\ll\tau_H$) and lie along the same scale as *h* and *E*, with fitness depending on the distance between (*h*+*b*) and *E*(*t*) (Sect. 4, Eq. (8)). Variation in behaviour is driven by $\lambda_B$, which may originate from molecular stochastic processes (Faisal et al.

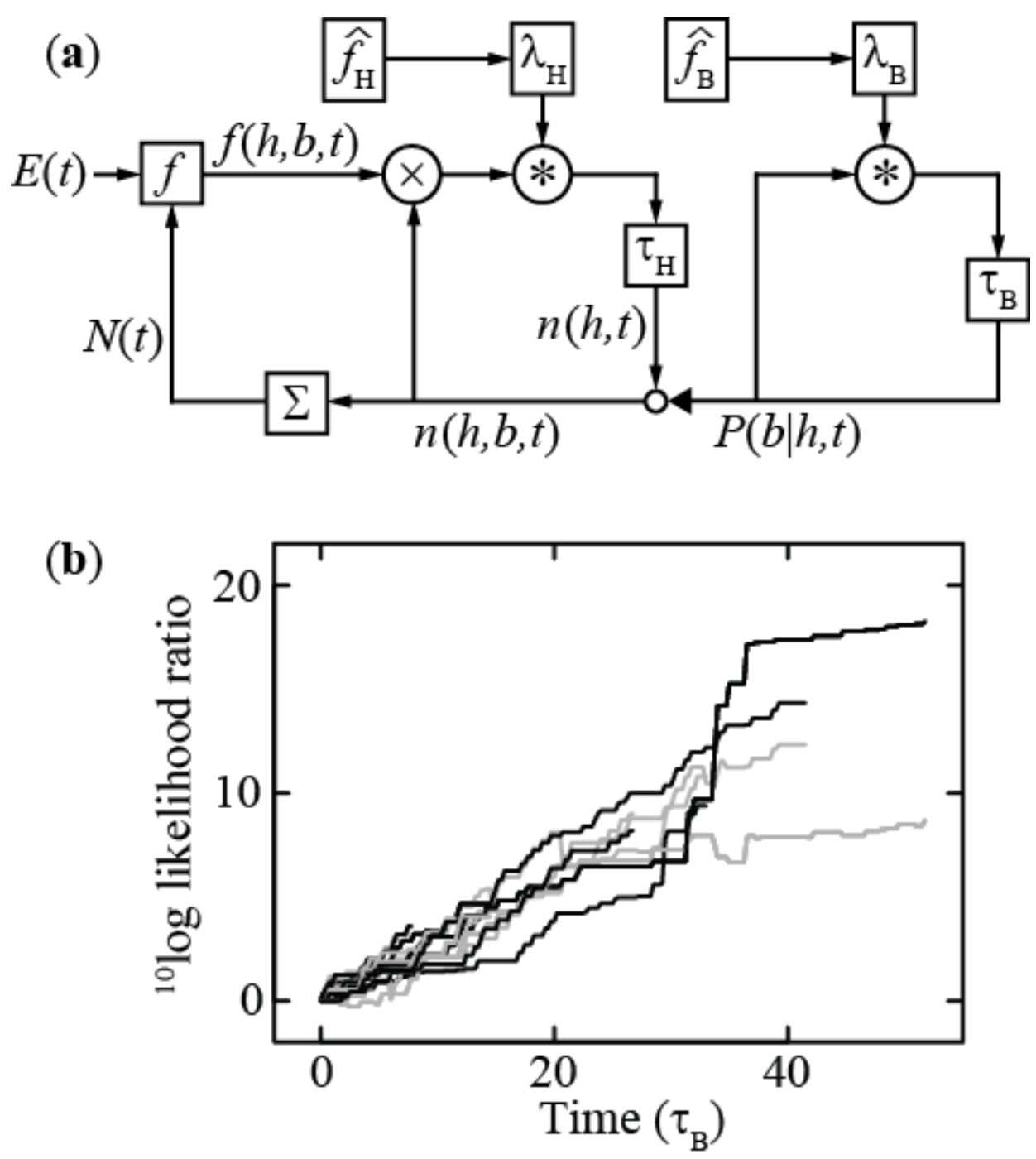

**Fig. 3** Behavioural extension of active causation. (**a**) Behaviour *b* adapts during the lifetime of an organism, in addition to the adaptation of *h* on an evolutionary timescale. The probability density of *b* for each *h*, $P(b|h,t)$ is varied through $\lambda_B(\hat{f}_B)$, with $\hat{f}_B(h,b,t)$ fitness estimated by the organism; $\tau_B$ is the typical lifetime of a specific behaviour, with for the calculations defaults $\tau_B$=4 and $\tau_H$=100 (in time step units). (**b**) Likelihood ratios of *b*-trajectories $\Pi_{\Delta b}P(\Delta b|\ \lambda_{B,modulated})/\ P(\Delta b|\ \lambda_{B,fixed})$ typically rise to high values. Shown are observed (grey) and expected (black) ratios of all *b*-trajectories in a 5$\tau_H$ interval of three different *h*-trajectories. The increasing likelihood ratios show that *b*-trajectories gradually become nearly completely determined by the goal function $\hat{f}_B$, even if this only involves random behavioural changes. See text for explanation and Sect. 4.5 for details

2008). The probability distribution of *b* given *h* and *t* determines which behaviours are present in the population. Populations that modulate $\lambda_B$ using a fitness estimate $\hat{f}_B$ outperform populations with fixed $\lambda_B$ (Sect. 4, Fig. 6), and the scheme is therefore evolvable. The *b*-trajectories (consisting of consecutive behaviours) of particular organisms (model shown in Fig. 5, Sect. 4) are only compatible with modulated $\lambda_B$ (Fig. 3b). This shows that the organism determines the trajectory in a decisive way, through $\hat{f}_B$, even if the details of the behaviour are unpredictable. Because, similarly to $\hat{f}_H$, also $\hat{f}_B$ is under evolutionary pressure to align itself with *f*, high $\hat{f}_B$ represents a goal for the organism. The feedback scheme also produces active causation as before, but now on a behavioural timescale. This produces a form of agency in the organism, with some behavioural freedom produced because the unpredictability of behaviour associated with stochasticity is modulated, and thus partly controlled, by the organism itself, through $\hat{f}_B$. The organism therefore actively pursues genuine goals on a behavioural timescale similarly as the organismal lineage does on an evolutionary timescale. Although the overall goal is high $\hat{f}_B$ (and, by implication, high *f*), this will in practice consist of a package of sub-goals, such as finding food and avoiding danger, all expected to contribute to the overall goal.

Organisms with advanced nervous systems can go one step further by first simulating the expected fitness consequences ( $\hat{f}_M$ ) of behavioural alternatives *m* before choosing a specific behaviour *b* (model shown in Fig. 7, Sect. 4). Again, populations that actively modulate $\lambda_M$ outperform populations that do not (Sect. 4, Fig. 8), and display active causation (Sect. 4, model in Fig. 9 and trajectories in Fig. 10). Because the consequences of potential behaviours are evaluated beforehand, the behaviour can be viewed as purposeful rather than just goal-directed.

Finally, organisms may communicate in order to coordinate their behaviour and thereby increase fitness. One possibility is to communicate through stereotypical signs that the other organism uses as auxiliary input to its $\hat{f}_M$, with its behaviour influenced by the resulting change in the value of $\hat{f}_M$. Communication in insects, such as by the honey-bee's dance, is presumably of this type. No intrinsic meaning is transferred, the communicated signs merely utilize the other organism's $\hat{f}_M$, with its implied intrinsic meaning.

A more sophisticated possibility is letting communication modify the very form of $\hat{f}_{\mathrm{M}}$. Changing the form of $\hat{f}_{\mathrm{M}}$ implies changing the intrinsic meaning it embodies. This type of communication then exchanges, in effect, intrinsic meaning between organisms, entangling their active causation. An example is presumably the nonverbal dialogue accompanying mother-offspring bonds in mammals, and particularly in humans (Trevarthen and Aitken 2001). Intrinsic meaning is a real phenomenon, on a par with any other aspect of the physical world, even if it only occurs in specialized, living forms of matter. In most organisms, it is presumably present only implicitly, distributed over many variables. However, once it becomes part of a dialogue, it needs to be made explicit and coded into behaviour, and the other organism needs to perceive and decode it, and assimilate it into its own intrinsic meaning. I conjecture that the process of extracting and assimilating intrinsic meaning during dialogue manifests itself as consciousness. The intuition here is that transfer of a real physical phenomenon (intrinsic meaning) gives rise to another real physical phenomenon (consciousness), somewhat analogous to how transfer of charge gives rise to a magnetic field even without an actual current (as in Maxwell's displacement current). Initially occurring only between two individuals, it may gradually extend, through development and learning, to more complex forms (Reddy 2003), such as nonverbal dialogue with another individual about objects (Tomasello and Carpenter 2007), nonverbal dialogue with internalized versions of others and the shared world, and with an internalized version of the self (Mead 1934). The scope for exchanging intrinsic meaning obviously expands considerably once the transition from nonverbal to verbal dialogue is made (van Hateren 2014). Finally, the combination of consciousness and active, goal-directed behaviour and dialogue can provide a physical mechanism for free will.

## 3 Discussion

The models presented here show how life can evolve two phenomena that are absent from the non-living world, but appear to be present in all current life-forms (van Hateren 2013). First, it evolved an active form of causation, enabling organismal lineages and organisms to initiate and pursue trajectories of heredity, behaviour, planned behaviour, and dialogue. Second, life evolved intrinsic meaning as embodied in the way organisms estimate their own fitness. Serving only as proofs of concept, the models ignore many aspects of existing life that are known to be important in evolution (Pigliucci and Müller 2010), development, and learning (Dehaene and Changeux 2000). Analysing the consequences and validity of the ideas presented here for more realistic models, and performing experimental tests, is evidently an extensive research program. However, the concepts of active causation and intrinsic meaning are based on such fundamental considerations that confidence in their robustness is justified.

The arguments presented here suggest that agency and intrinsic meaning are emergent properties when hereditary or behavioural variability is based on an organism's estimation of its own fitness. With respect to hereditary variability, this is closely related to second-order selection, selection for properties enhancing evolutionary potential (see e.g. Kauffman 1993). However, when this selection works for example on mutator alleles (Taddei et al. 1997), with the hindsight of natural selection, the causation is merely passive. In the extended scheme of Fig. 1c, causation is active because it involves a predictive estimate of fitness, made by the lineage's organisms themselves and utilized in a feedback loop that transforms the randomness of hereditary change into goal-driven hereditary trajectories. Similarly, a behavioural feedback loop incorporating self-estimated fitness leads to goal-driven behavioural trajectories of an organism. Although the simulations used an estimate $\hat{f}$ identical to *f*, the scheme is still fundamentally different from one implementing passive second-order selection. In passive second-order selection, the mutation rate (as set by mutator alleles) is established by natural selection, thus after the fact (based on a realized fitness outcome, not a predictive one). In contrast, the fitness $\hat{f}$ in the scheme of Fig. 1c is an estimate of a predictive, probabilistic variable, *f*, and it is therefore also predictive and probabilistic, with “as if” foresight. The foresight is not real, of course, because that would conflict with causality. Rather, it is based on implicit statistical knowledge the organismal lineage has acquired through previous natural selection (or, for behaviour, through previous development and learning as guided by criteria acquired through previous natural selection).

In this article it was assumed that organisms evolved reasonable estimates $\hat{f}$ of their true fitness *f*, through regular evolutionary mechanisms. For the toy model it was assumed that $\hat{f}$ could be approximated by *f* (as would indeed be easy for the toy organisms given the simplicity of their fitness function), but it is likely that also organisms in the wild can readily evolve practical ways to estimate and internalize *f*. Nevertheless, it is a deep and interesting question if there are optimal ways for organisms to make estimates of *f*, if such estimates are subject to theoretical bounds, and to what extent actual organisms reach those bounds. An organism can obtain high fitness by two basic factors, self-maintenance and reproduction. When self-maintenance is too weak, the organism will die before it can reproduce. When reproduction is too weak, the organism will be the last of its lineage. Self-maintenance is related to homeostasis and autopoiesis, and, according to the good regulator theorem (Conant and Ashby 1970), a good regulator (or homeostat) has to model its environment and how it is affected by it. It is clear that such a model must, implicitly, play an important role in estimating fitness as well. Thus internalizing *f* as a model $\hat{f}$ that can predict the environment, as part of its capabilities, is consistent with regulating well. Indeed, there is a venerable research tradition in neuroscience and machine learning that focusses on making such inferences by utilizing Bayesian methods. A recent formulation in terms of minimizing free energy (Friston and Stephan 2007; Friston 2010) unifies earlier approaches with respect to sensory perception (such as predictive coding and maximization of mutual information) with action (behaviour) as a means of increasing perceptual predictability. An interesting possibility is that the feedback schemes presented here might be used as effective mechanisms for optimising variational free energy through selection at evolutionary or somatic timescales. This would require that fitness can be specified by the states that constitute an organism's goals. Alternatively, minimizing free energy may then be seen as an optimal strategy by which an organism could estimate its own fitness, an upper bound that can be compared with the (possibly contingent) strategy actually realized in the biological system. A caveat here is that informational approaches, such as the free energy principle, generally assume ergodicity, whereas the current schemes presumably produce organismal lineages and organisms with non-ergodic properties: their state space changes over time as organisms innovate by reconstructing themselves through genetic, physiological, and social mechanisms. Reconstructing $\hat{f}$ by social bonding is conjectured (Sect. 2.2) to be associated with consciousness.

The word “active” is quite general, and it is used here in a more restrictive sense than in related work where the term is also used. In particular, approaches stressing homeostasis and autopoiesis (Di Paolo 2005) view an organism that is autonomous and adaptive, and that appears to act on its own behalf (Kauffman 2003), as having agency and being active. Similarly, in the free energy approach, when an organism maintains internal homeostasis by acting on its environment and modifying its sensory inputs, this is called active inference (essentially because it involves the part of the free energy formalism that describes action). However, systems described in these approaches will usually involve only regular deterministic and stochastic causation, which is called passive here (see also Sect. 5). Only the entangled form of causation described in Sect. 2, produced by a feedback loop involving estimated fitness and modulated stochastic causation, is called active here, and it is only that specific feedback loop, or more complex versions of it, that is able to generate genuine meaning (as an emergent property evolved by natural selection).

The models presented here took inspiration from so-called two-stage models of free will, proposed by many researchers going back at least to William James (reviewed in Doyle 2013). In these models, a first stage generates random behavioural options, one of which is subsequently selected by a deterministic second stage. Such models and variants of them (Tse 2013) are themselves inspired by Darwin's evolution theory, and conform to the model of Fig. 1a. They are therefore restricted to passive causation and do not embody intrinsic meaning, essentially because they do not let the randomness $\lambda$ be a function of a self-estimated fitness $\hat{f}$, in a feedback loop and with $\hat{f}$ under evolutionary pressure to align itself with the true fitness *f*. A similar restriction applies to the theory of neuronal group selection (Edelman 1993) and to work in the functionalist tradition (e.g., Pugh 1976), which also conform to or rely on the basic Darwinian scheme, and therefore lack the active causation and intrinsic meaning produced by $\hat{f}$ in the extended behavioural scheme (Fig. 3, and Sect. 4, Fig. 7).

Complex feedbacks between evolution and development, and between how organisms shape their own environment and how that influences their subsequent evolution (Pigliucci and Müller 2010), are occasionally presented in wording that suggests goal-directedness. However, without a specific term $\lambda(\hat{f})$ modulating randomness, such feedback schemes are elaborations of the basic Darwinian scheme of Fig. 1a, containing only passive causation. The goal-directedness and agency of the organism they suggest are therefore not genuine, but merely "as if".

It is sometimes assumed that any observed stochasticity in the world is not real, but only apparent, caused by a lack of complete knowledge of the underlying causes of the observables. The models and arguments presented in this article are agnostic with respect to the source and nature of the stochasticity required, and are therefore not affected by any such assumptions. Nevertheless, at a deeper level it might be argued that, if stochasticity is only apparent, the freedom provided by active causation is only apparent as well. I believe that such a view is likely to be mistaken for several reasons. First, the "as if"-stochasticity argument ignores quantum indeterminacy, which appears to be a genuine source of randomness. Second, the argument is often based on the theoretical assumption that current physical theories have unlimited validity, whereas these theories are known to be incomplete, and it is not clear how far they can be extrapolated beyond the range where they have been tested. Finally, in a classical, Newtonian universe, even Laplace's demon gets confused, within a tiny fraction of a second, if the whereabouts of merely a single electron are unknown (Berry 1988). Arguing that microscopic fluctuations cannot show at macroscopic scales because they average out is not valid either, because nonlinearities can easily amplify even the smallest indeterminacies to prominence (Berry 1988; Laskar and Gastineau 2009).

The relationship between intrinsic meaning and fitness, as well as fitness itself, is highly complex in advanced organisms. Fitness extends to related organisms and organisms with shared interests, and depends on feedbacks through a malleable environment. The dual inheritance system in humans (molecular heredity alongside cultural inheritance) further complicates matters (Boyd and Richerson 1985; Boyd et al. 2011), as cultural flexibility makes human fitness a moving target. Fully reducing human meaning to fitness is therefore almost certainly intractable (Anderson 1972), although it may be possible to derive broad generalities relating the two. This article does indeed not intend to advocate such a reduction, but rather to show how the phenomena of meaning and goal-directed initiative can be understood, at least in principle, from existing scientific knowledge.

## 4 Methods

### 4.1 Basic model (Fig. 1a)

For simplicity, $n(h,t)$ is taken to be continuous. The dynamics of the primary feedback loop $I$ is governed by a first-order low-pass filter with time constant $\tau_H$

$$\tau_H \frac{dn(h,t)}{dt} = \Lambda * (f(h,t)n(h,t)) - n(h,t), \tag{1}$$

where the asterisk denotes convolution along the $h$-coordinate. $\Lambda$ is a normalized Gaussian with coordinate $h$ and standard deviation $\lambda_H$. Fitness is given by

$$f(h,t) = f_{max} \exp(-(h - E(t))^2 / 2\sigma_E^2) \exp(-N(t)/K), \tag{2}$$

with $N(t) = \sum_h n(h,t)$, $f_{max}$ the maximum fitness, $\sigma_E$ determining the width of the fitness function, and $K$ the carrying capacity of the environment (for simplicity not taken as a hard limit, but one that produces a gradually decreasing fitness). For all calculations $\tau_H$=100, $f_{max}$=3, $\sigma_E$=5, and $K$=10000 are used (units in discrete time steps and units of $h$). The value of $K$ is only nominal, because continuous $n$ implies an effectively infinite population size. The environmental variable $E(t)$ has units of $h$ and is taken as filtered Gaussian white noise, using as filter a normalized sum of low-pass filters, with pulse response (for $t \geq 0$; for $t<0$ $h(t)=0$) given by

$$h(t) = \sum_{i=1}^{k} \frac{a_i}{\tau_i} e^{-t/\tau_i}, \tag{3}$$

with

$$\tau_i = r^{i-1}\tau_z \quad (4)$$

$$a_i = \tau_i^q \Big/ \sum_{i=1}^{k} \tau_i^q \ . \quad (5)$$

The noise is generated with a standard deviation of 1000 per time step, and the filter parameters are $\tau_z$=4, $r$=4, $k$=6, and $q$=1. Figure 4a shows an example of $E(t)$ (for computational convenience shifted to positive values only), and Fig. 4b its power spectral density. The latter follows approximately a power law, with a slope of about -1.8. The power-law behaviour reverts to a flat spectrum for frequencies much smaller than $1/\tau_6$ (=1/4096; not shown in the figure), which means that the power spectral density is integrable. For the results presented in this article, the precise form of $E(t)$, for example as steered by $q$, is not critical, as long as it contains sufficient power at low frequencies such that adaptation of $h$ with a relaxation time $\tau_H$ increases fitness, and at the same time sufficient power at high frequencies such that adaptation of $b$ with a relaxation time $\tau_B$ increases fitness as well.

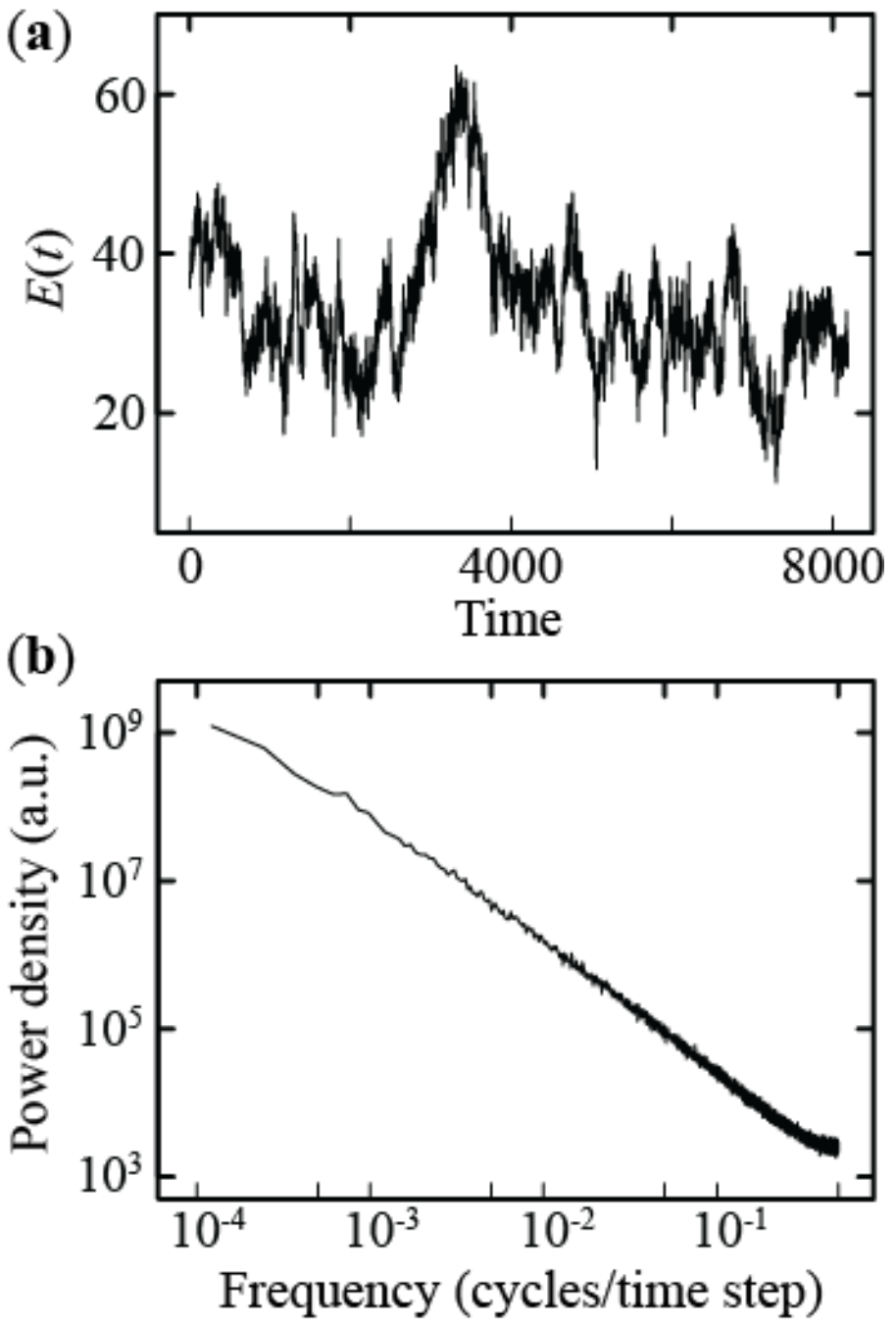

**Fig. 4** Properties of environmental variable *E*. (**a**) Example of *E*(*t*) for 8192 time steps. (**b**) Power spectral density of *E*(*t*), periodogram average of the spectra of 100 realizations of *E*(*t*) of length 8192 time steps each; a.u.: arbitrary units

In the calculations, heredity $h$ is discretized to integer values, and calculations are started with $K$ organisms (either in one population or evenly divided between two populations) of type $h=E(0)$ convolved with $\Lambda$. Low-pass filters (here and below) are implemented as autoregressive filters $y_i=f_1 y_{i-1}+f_2 x_i$, with $x$ input, $y$ output, and the coefficients given by $f_1=\exp(-1/\tau)$ and $f_2=1-f_1$ (van Hateren 2008). Models are implemented in R, with calls to Fortran routines at places critical to keep computation times manageable. Sources are available at https://sites.google.com/site/jhvanhateren/ or upon request from the author.

Simulations show that for basic Darwinian evolution populations can adapt to $E(t)$, provided the mutation rate $\lambda_H$ is chosen suitably. However, if $\lambda_H$ is chosen too small, fast changes in $E(t)$, which occur unpredictably, strongly reduce the population size because the new types $h$ required are not generated fast enough. If $\lambda_H$ is chosen too large, types $h$ that are well adapted to $E(t)$ mutate away too quickly, again strongly reducing the population size. For the $E(t)$ used here, a value of $\lambda_H$=2 was found to be close to optimal to sustain a viable population, and this value is therefore used as default for simulations with fixed $\lambda_H$.

The overall structure of the model has some resemblance to the quasispecies theory of Eigen and Schuster (1977) and other approaches to evolutionary dynamics (Page and Nowak 2002), but is different in details. In particular the way heredity and changes in heredity are handled and the relationship between $E(t)$, $h$, and fitness are highly simplified, with mutations, though undirected, always on the same scale as $h$ and $E(t)$ and therefore never completely lethal (but possibly of low fitness).

4.2 Extended model (Fig. 1c, d)

As a reasonable choice an inverse relationship between $\hat{f}_H$ and $\lambda_H$ is used with empirically chosen parameter values that proved to work well. Quite likely, other relationships and different parameter values will work even better, but as the aim of the model computations is only to provide a proof of concept, and the model is a mere toy model, this was not investigated further. The default value of $\lambda_H$=2.0 is multiplied by a factor as

$$\lambda_H(\hat{f}_H) = 2.0 \times 1.8/(\hat{f}_H + 0.1)\ , \quad (6)$$

which is a factor fairly close to one for values of the fitness $\hat{f}_{\mathrm{H}}$ in the typical high fitness range of 1.3-2.3, and fairly large (limited by the 0.1 in the denominator) for values of $\hat{f}_{\mathrm{H}}$ close to zero. For large values of $\lambda_{\mathrm{H}}$ some of the resulting *h* values can fall beyond the range of *h* considered, and those organisms are removed from the computation (becoming extinct). For simplicity, the model calculations use an estimated fitness $\hat{f}_{\mathrm{H}}$ identical to the true fitness *f*, with similar assumptions for $\hat{f}_{\mathrm{B}}$ and $\hat{f}_{\mathrm{M}}$ below. However, the model calculations are not very sensitive to this assumption: deliberately introducing errors in $\hat{f}_{\mathrm{H}}$ still produces results similar to those of Fig. 1d (in Eq. (2), with *f* as used for $\hat{f}_{\mathrm{H}}$ in Eq. (6), 10% increases or decreases of $f_{max}$ or $\sigma_{\mathrm{E}}$, or shifting $h$-$E(t)$ by one unit, corresponding to errors in observing *h* or $E(t)$).

Evolvability of the extended model (Fig. 1d) was evaluated as follows. Two populations $n_{\lambda,\mathrm{modulated}}$ (according to the extended model) and $n_{\lambda,\mathrm{fixed}}$ (according to the basic model) are interacting and competing by using as *N* in Eq. (2) $N$=$N_{\lambda,\mathrm{modulated}}$+$N_{\lambda,\mathrm{fixed}}$, with $N_{\lambda,\mathrm{modulated}}=\Sigma_h n_{\lambda,\mathrm{modulated}}$ and $N_{\lambda,\mathrm{fixed}}=\Sigma_h n_{\lambda,\mathrm{fixed}}$. The figure shows the results of 100 such computations, with $\lambda_{\mathrm{H,modulated}}$ always according to Eq. (6) and $\lambda_{\mathrm{H,fixed}}$ systematically varied from 0.5 to 3.5. In all cases the extended model outperforms the basic model, yielding a ratio $N_{\lambda,\mathrm{fixed}}/ N_{\lambda,\mathrm{modulated}}$ that fluctuates, but gradually declines with time.

4.3 Trajectory version of the extended model (Fig. 2a)

Figure 2a is the equivalent of Fig. 1c for individual *h*-trajectories. Collections of *h*-trajectories produce results consistent with the population calculations (as is also the case for the *b*-trajectories of the *b*- and *m*-models below). Trajectories are calculated alongside with the population $n(h,t)$, from which $N(t)$ is used for calculating fitness as needed for the trajectory calculations. Fitness determines the number of offspring by a Poisson random generator ($R_{\mathrm{poisson}}$, with parameter *f*), yielding 0, 1, 2, or more offspring (split arrows at $\Delta n$ in the figure). With $\Delta n$=0 the trajectory (lineage) becomes extinct, and is removed from the computation. With $\Delta n$=1 the trajectory just continues (with a new step in heredity $\Delta h$ , thus *h* becomes $h+\Delta h$, where the new *h* lasts for a time $\Delta t_{\mathrm{H}}$ before new offspring is generated). With $\Delta n$>1 the trajectory splits in two or more, each of which trajectories is subsequently treated separately. The steps in heredity are generated by a normal (Gaussian) random generator ($R_{\mathrm{normal}}$, with parameter $\lambda_{\mathrm{H}}$), and the lifetime $\Delta t_{\mathrm{H}}$ by an exponential random generator ($R_{\exp}$, with parameter $\tau_{\mathrm{H}}$). Such a random process is the equivalent of the first-order low-pass filter in the population model (symbolized by $\tau_{\mathrm{H}}$ in Fig. 1c).

4.4 Likelihood ratio of *h*-trajectories (Fig. 2b)

The likelihood ratio as a function of time *t* is calculated as the product $\Pi_{\Delta h} P(\Delta h|\ \lambda_{\mathrm{H,modulated}})/\ P(\Delta h|\ \lambda_{\mathrm{H,fixed}})$ over all steps $\Delta h$ of a particular trajectory up to time *t*. Whereas $P(\Delta h|\lambda_{\mathrm{H,modulated}})$ is given by a normal distribution with standard deviation $\lambda$ according to Eq. (6), the choice of $\lambda_{\mathrm{H,fixed}}$ is somewhat arbitrary, but influences the likelihood ratio. Therefore the $\lambda_{\mathrm{H,fixed}}$ producing the minimum mean end value of the likelihood ratios of all *h*-trajectories computed in a simulation was determined numerically, and that value of $\lambda_{\mathrm{H,fixed}}$ is used for the figure.

Because the mathematical forms of $P(\Delta h|\ \lambda_{\mathrm{H,modulated}})$ and $P(\Delta h|\ \lambda_{\mathrm{H,fixed}})$ are known (both Gaussians), it is possible to compute an expected likelihood ratio function $B(t)$. This only requires $\lambda_{\mathrm{H,modulated}}$ as a function of time rather than the series of random realizations of $\Delta h$ produced by the $R_{\mathrm{normal}}$ of Fig. 2a. We write $\lambda_0$ for $\lambda_{\mathrm{H,fixed}}$ and $\lambda_i$, $i$=1..*L* for the successive $\lambda_{\mathrm{H,modulated}}$ of *L* organisms in a trajectory, and we need to integrate over $P(\Delta h_1,..,\Delta h_L|\lambda_1,..,\lambda_L)$ if the $\Delta h_i$ are actually produced by the extended scheme:

$$\begin{aligned}\langle \log(B) \rangle &= \int d\Delta h_1 P(\Delta h_1 \mid \lambda_1) \ldots \int d\Delta h_L P(\Delta h_L \mid \lambda_L) \log\left( \prod_{i=1}^{L} \left( P(\Delta h_i \mid \lambda_i) / P(\Delta h_i \mid \lambda_0) \right) \right) \\ &= \int d\Delta h_1 P(\Delta h_1 \mid \lambda_1) \ldots \int d\Delta h_L P(\Delta h_L \mid \lambda_L) \left( \sum_{i=1}^{L} \log\left( P(\Delta h_i \mid \lambda_i) / P(\Delta h_i \mid \lambda_0) \right) \right) \\ &= \sum_{i=1}^{L} \int d\Delta h_i \frac{1}{\sqrt{2\pi}\lambda_i} \exp(-(\Delta h_i)^2 / 2\lambda_i^2) \left[ \log \frac{\lambda_0}{\lambda_i} - (\Delta h_i)^2 \left( \frac{1}{2\lambda_i^2} - \frac{1}{2\lambda_0^2} \right) \right] \\ &= \sum_{i=1}^{L} \left( \log \frac{\lambda_0}{\lambda_i} - \frac{1}{2} + \frac{1}{2} \frac{\lambda_i^2}{\lambda_0^2} \right). \end{aligned} \tag{7}$$

The result gives the expected natural logarithm of the likelihood ratio, which is converted to the logarithm base 10 for the figure.

4.5 Extended model with behaviour (Fig. 3a, b)

Fitness is now given by

$$f(h,t) = f_{\max} \exp(-(h + b - E(t))^2 / 2\sigma_E^2) \exp(-N(t)/K), \tag{8}$$

with the probability of $b$ for each $h$ given by $P(b|h,t)$. The $h$-loop on the left of Fig. 3a uses a fitness estimate $\hat{f}_H(h,b,t)$, but eventually produces $n(h,t)$ as summed over $b$. With $P(b|h,t)$ this is subsequently converted to $n(h,b,t)= n(h,t)P(b|h,t)$. Typical lifetime of a behaviour $b$ is $\tau_B$ (defaulted to 4), and new $b$ are generated by convolution with a normalized Gaussian with standard deviation $\lambda_B$. Calculations are started with the population at $b$=0 convolved with this Gaussian. For the modulated scheme $\lambda_B$ is driven by a fitness estimate $\hat{f}_B(h,b,t)$ according to

$$\lambda_B(\hat{f}_B) = 0.5 \times 1.8 / (\hat{f}_B + 0.1), \tag{9}$$

again determined empirically as a suitable choice. The allowed range of $b$ is taken as $-b_{max}..b_{max}$. If $b_{max}$ is small compared with $\sigma_E$ and $\lambda_H$, behaviour has little effect on fitness, and all adaptation occurs by changes in $h$. If $b_{max}$ is large, all adaptation can be done quickly by behavioural changes, and the distribution over $h$ will tend to change little. A choice of $b_{max}$=10 produces a well balanced combination of $h$ and $b$ adaptations, and is used in the calculations. When convolution with $\lambda_B$ produces values of $b$ beyond the allowed range, values are replaced as in circular convolution (continuing on the other side of the range). Qualitatively similar results are obtained when folding $b$ back as if reflected at the boundary of the allowed range.

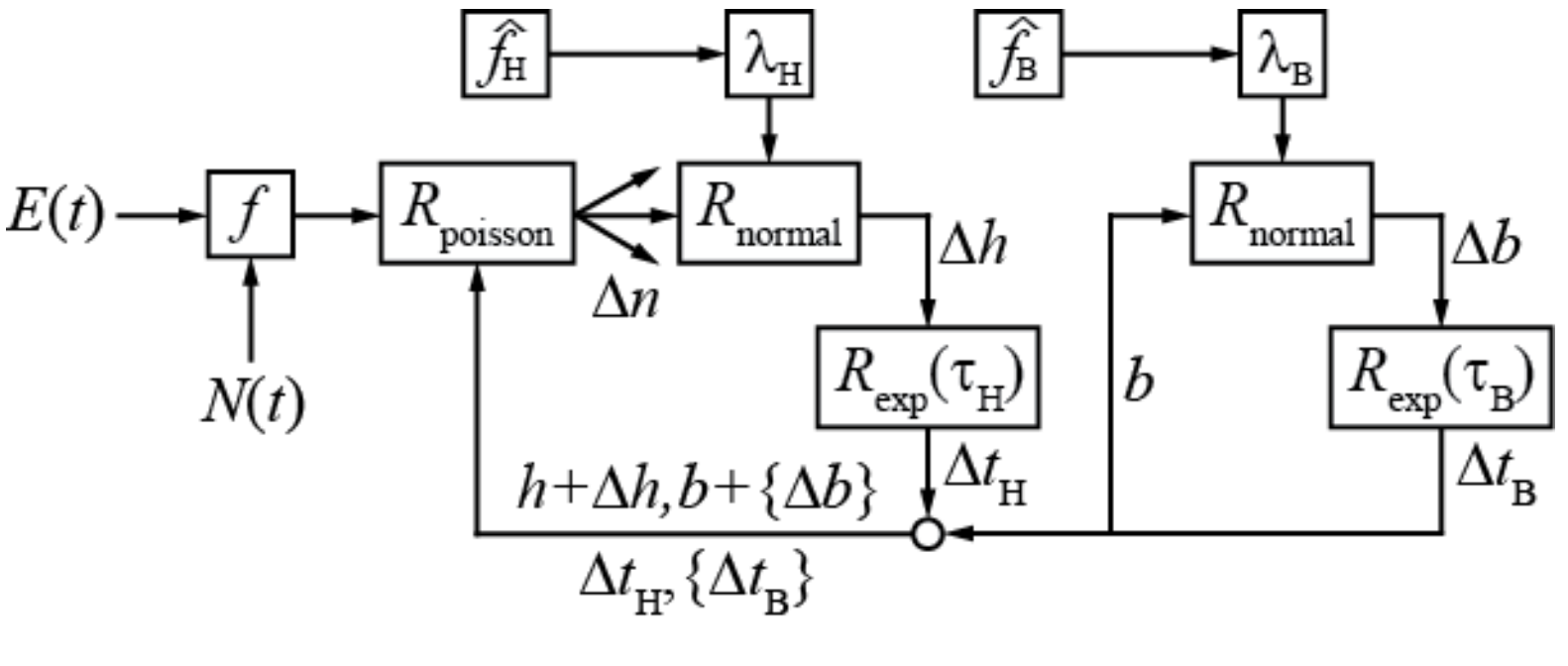

**Fig. 5** Equivalent of Fig. 3a (which provides $N$) for computing single $h$- and $b$-trajectories, with a $b$-trajectory the trajectory (through behavioural space) of consecutive behaviours of a particular organism. $R$ are random generators. Details are given in Sect. 4.5

For computing the likelihood ratios of $b$-trajectories the same methods were used as described above for Fig. 2b. The equivalent of Fig. 3a for trajectories (Fig. 5) shows how $b$-trajectories are generated within each $h$-trajectory. For simplicity, the trajectories are generated asynchronously, as is implicitly the case for the population model of Fig. 3a as well. This implies that a new $h$-trajectory may continue, for a relatively short time ($\tau_B$=4 relative to $\tau_H$=100), with the previous $b$ value, which subsequently can diffuse away quickly to other $b$ values.

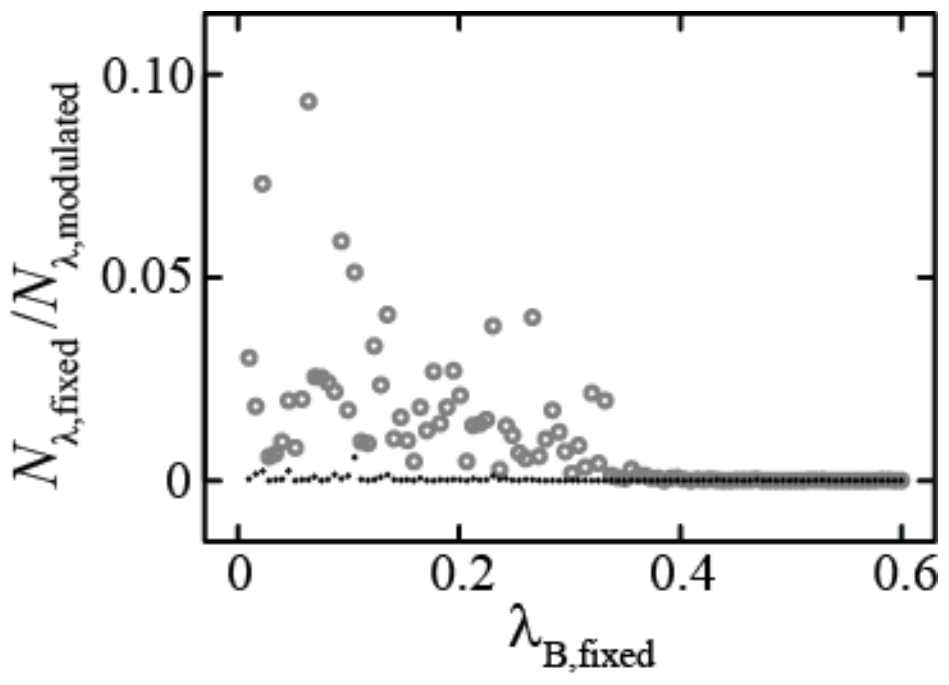

**Fig. 6** Ratios of population numbers $N$ of 100 pairs of competing populations, one always following the model of Fig. 3a (with $\lambda_{B,\text{modulated}}$) and the other the corresponding model with fixed $\lambda_B$, over a range of $\lambda_{B,\text{fixed}}$. Results after $30\tau_H$ (open grey symbols) and $60\tau_H$ (black dots). The results show that the extended model is evolvable also for behaviour. Details are given in Sec. 4.6

4.6 Evolvability of the modulated behavioural model (Fig. 6)

Two populations $n_{\lambda,\text{modulated}}$ (according to the model with modulated $\lambda_B$ and modulated $\lambda_H$) and $n_{\lambda,\text{fixed}}$ (according to the model with fixed $\lambda_B$, but modulated $\lambda_H$) are interacting and competing by using as $N$ in Eq. (8) $N=N_{\lambda,\text{modulated}}+N_{\lambda,\text{fixed}}$. The figure shows the results of 100 such computations, with $\lambda_{B,\text{modulated}}$ always according to Eq. (9) and $\lambda_{B,\text{fixed}}$ systematically varied from 0.01 to 0.6. In all cases the modulated model outperforms the fixed model. The limit for small $\lambda_{B,\text{fixed}}$ is in fact identical to the model without behavioural adaptations (Fig. 1c), which indicates that the ability to produce behavioural adaptations is evolvable (depending on the fitness costs of behaviour, such as more resources needed for operation, longer maturation times, and so on).

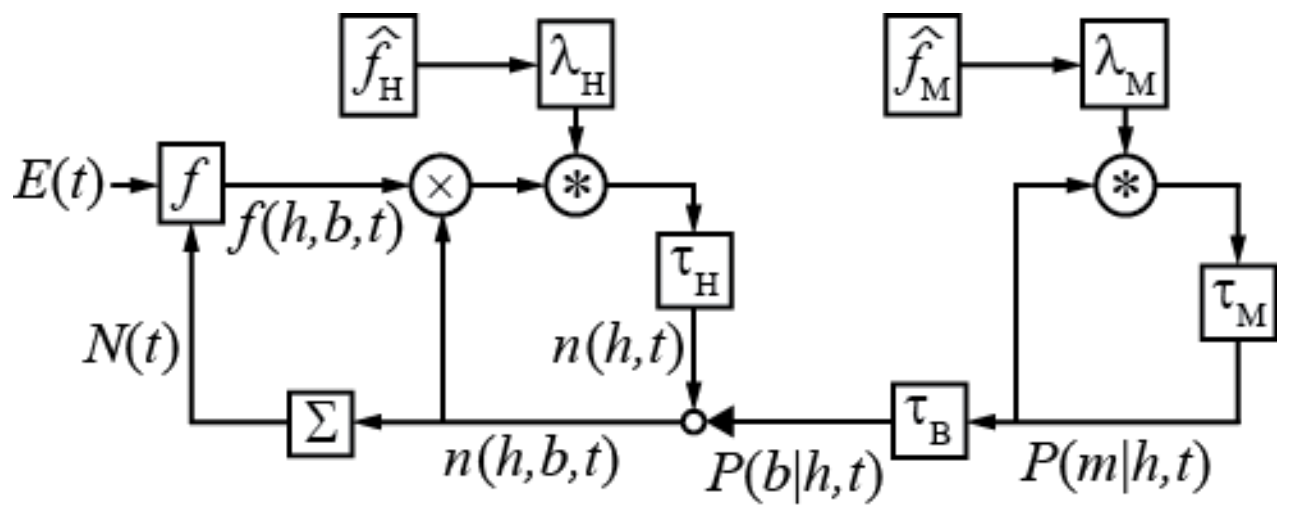

**Fig. 7** Motor planning extension of active causation. The probability density of motor plan $m$ for each $h$ is varied through $\lambda_M(\hat{f}_M)$, with $\hat{f}_M(h,m,t)$ fitness estimated by the organism. $P(b|h,t)$ is a low-pass filtered version (with $\tau_B>\tau_M$) of $P(m|h,t)$. Because the motor planning is simulated, and can be implemented in parallel in the nervous system, it can be faster than actual behaviour, and the effect of a specific behaviour can be simulated, based on $\hat{f}_M(h,m,t)$, before it is actually executed. For the calculations, defaults are $\tau_M=1$, $\tau_B=4$, and $\tau_H=100$

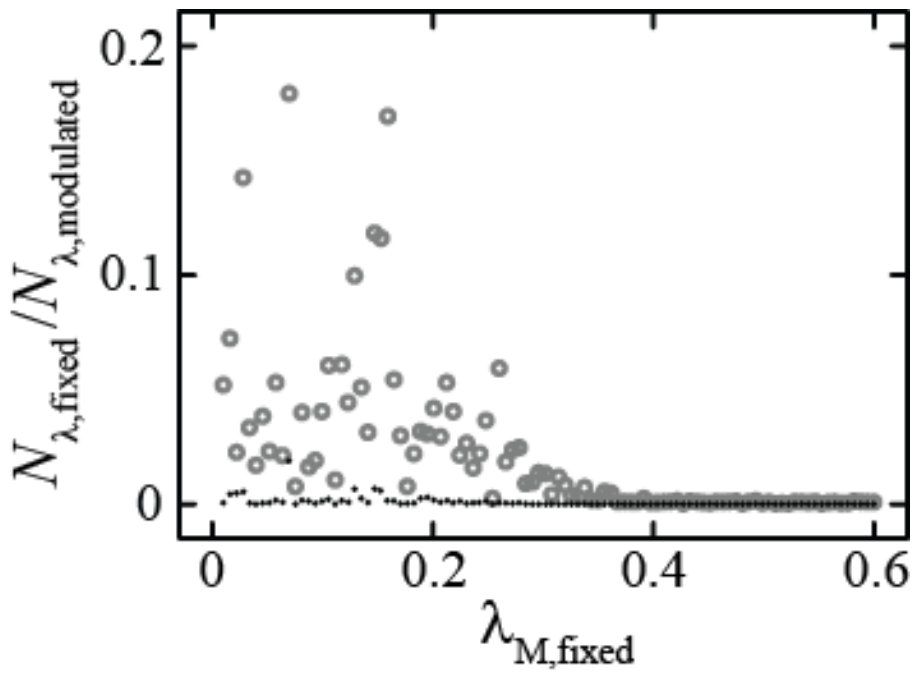

**Fig. 8** Ratios of population numbers $N$ of 100 pairs of competing populations, one always following the model of Fig. 7 (with $\lambda_{M,\text{modulated}}$) and the other the corresponding model with fixed $\lambda_M$, over a range of $\lambda_{M,\text{fixed}}$. Results after $20\tau_H$ (open grey symbols) and $40\tau_H$ (black dots). The results show that the extended model is evolvable also for planned behaviour

4.7 Trajectory version of the motor planning model and *b*-trajectories (Figs. 9 and 10)

Figure 9 shows how first $P(m|h,t)$ is generated (in parallel, because there is no requirement to generate serial trajectories as all operations take place within individual organisms). $P(m|h,t)$ is subsequently converted to $P(b|h,t)$ through the behavioural low-pass filter $\tau_B$, and from $P(b|h,t)$ *b*-trajectories are generated within each *h*-trajectory. For computing the likelihood ratios of *b*-trajectories (Fig. 10) only the direct method $\Pi_{\Delta b}P(\Delta b|\ \lambda_{M,modulated})/\ P(\Delta b|\ \lambda_{M,fixed})$ could be used (not the expected $B(t)$ as in Eq. (7)), because there is no analytical expression for $P(m|h,t)$ available. Planned behaviour *m* is evaluated as $\hat{f}_M(h,m,t)$ and produces $\lambda_M$ according to Eq. (9) (with B replaced by M). The typical lifetime of *m* is $\tau_M$=1.

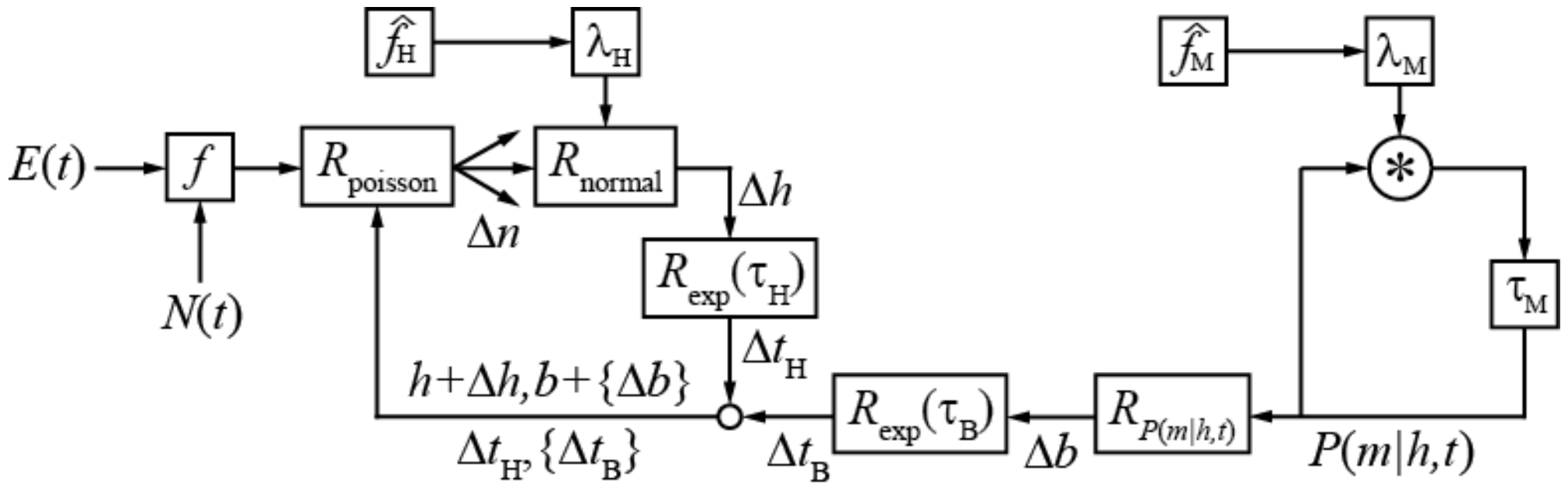

**Fig. 9** Equivalent of Fig. 7 (which provides *N*) for computing single *h*- and *b*-trajectories. *R* are random generators. Because *m* is computed and evaluated within single organisms, the model performs that part of the computation in parallel. The computation of *h* and *b* is serial as before. For details see Sect. 4.7

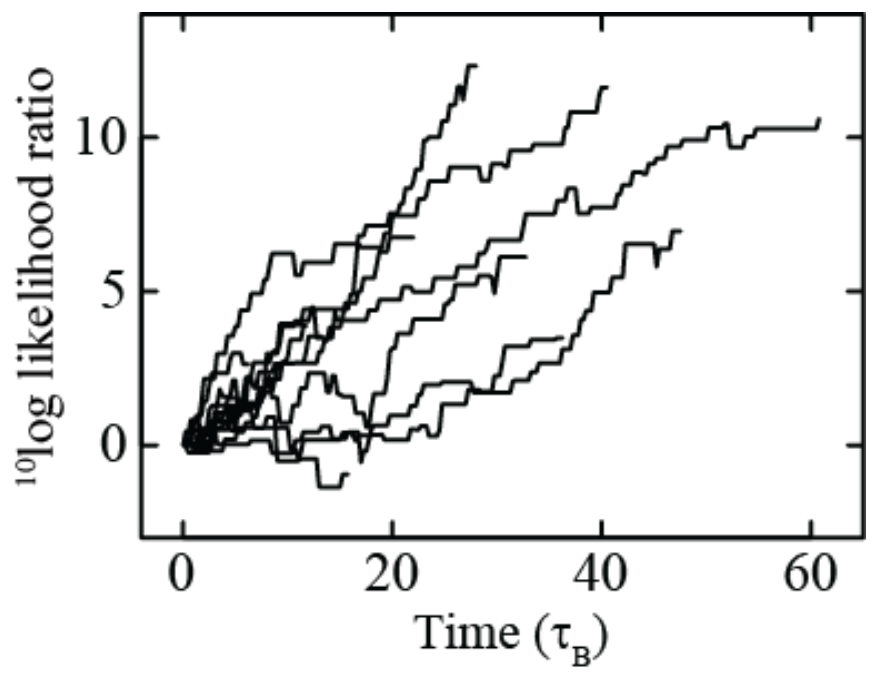

**Fig. 10** The likelihood ratios of *b*-trajectories, computed with the *m*-model of Fig. 9 as $\Pi_{\Delta b}P(\Delta b|\ \lambda_{M,modulated})/\ P(\Delta b|\ \lambda_{M,fixed})$, typically rise to high values. Shown are observed ratios of all *b*-trajectories in an $5\tau_H$ interval of three different *h*-trajectories. The increasing likelihood ratios show that *b*-trajectories gradually become nearly completely determined by the goal function $\hat{f}_M$, even if this only involves a neural simulation, made by the organism, of random behavioural changes. For details see Sect. 4.7

**Acknowledgements** I would like to thank the anonymous reviewers of the manuscript for thoughtful comments and suggestions.

## 5 Appendix: Terminology

Several of the terms used in this article, such as “meaning”, “active”, and “freedom” are veterans of natural language for describing human affairs, and they are therefore, as common in natural language, somewhat broad, vague, and ambiguous. In this section I will try to explain a bit more accurately than in the main text as to how I use them, in order to reduce their ambiguity within the present context.

The main point of this article is that an internalized fitness estimate, $\hat{f}$, is used in feedback loops as in Fig. 1c and Fig. 3a for modulating hereditary and behavioural variability. This estimate and its form represent what is designated by the term “intrinsic meaning”. This use of the word “meaning” is close to that given by the Oxford English Dictionary (OED online, accessed May 22, 2014) as [meaning, n.2, 1], for example as used in expressions like the meaning of an action or the meaning of life. It implies significance, import, value, and purpose. Such a broad spectrum of denotations is appropriate here, because using $\hat{f}$ is presumably a feature of all forms of life (van Hateren 2013), and it represents at the same time a goal-directedness (as in life striving for high fitness in order to survive and reproduce), a value in the form of high fitness, and, through the form of $\hat{f}$ (i.e., of which parameters it is a function, and in what form) which aspects of the environment and itself the organism takes into account for estimating its own fitness. Those aspects are the ones that represent significance and import to the organism. For most organisms the goal-directedness is merely an implicit drive in the direction of appropriate, though underdetermined end-points (together producing high fitness). However, in organisms where $\hat{f}$ becomes so complicated and flexible that it gives rise to planned behaviour (Figs. 7 and 9), consciousness, and language (van Hateren 2014), end-points may be produced by simulation or even conscious deliberation, and thus evolve into genuine purposes.

It is important to stress that “meaning” is used here not in the more narrow linguistic sense as in the meaning (signification) of a word [OED, meaning n.2, 2], and also not as a means for categorization and symbolic generalization. Nevertheless, it is likely that this more narrow sense can be understood as derivable from the broad sense, as the end result of an as yet not completely resolved chain of reasoning (sketched in van Hateren 2014). The term “intrinsic meaning” is also different from “value” in the sense of the variable defining a value-driven decision process: “intrinsic meaning” is not a single number, but includes the form of the estimated fitness function and the fact that there is an implicit drive towards high estimated fitness.

Apart from “intrinsic meaning”, the feedback loops of Fig. 1d and Fig. 3 also give rise to a special form of causation, because stochasticity is continually modulated by the goal-directedness of $\hat{f}$ and thus produces hereditary and behavioural trajectories that are unpredictable, but not random. This form of causation is designated by the term “active causation”, with the meaning of “active” close to that given in [OED, active, 2], i.e., capable of acting on something, originating action, spontaneous, voluntary. Active causation is contrasted with deterministic causation, which cannot originate anything new because systems subject to deterministic causation passively follow internal or external influences, like a clockwork. It is also contrasted with stochastic causation, which does originate new events and does so spontaneously (not caused by internal or external influences), but cannot be seen as producing action (because it lacks a goal-directedness) nor doing so voluntarily (because it lacks purpose). Both deterministic and stochastic causation are therefore designated by the term “passive causation” here, because either lacks several of the defining characteristics of “active”. The feedback loops of Fig. 1c and Fig. 3a entangle deterministic and stochastic causation in such a way that all defining characteristics of “active” are produced: there is the capability to originate and initiate in a spontaneous way (because of the stochasticity), but also in a way that can be seen as acting and performing action, because of the intrinsic goal-directedness produced by the feedback loops. For advanced organisms with reasoning and consciousness (van Hateren 2014), the acts can be deliberate and voluntary.

The ability to originate and perform action is a form of “freedom” for the organism, because its behaviour is neither dictated by a determined chain of causation, nor by the arbitrariness of randomness. This sense of freedom is close to [OED, freedom, 5], i.e., the power of self-determination (but without the mental or spiritual connotations), which can be seen as an absence of necessity [OED, necessity, 2], i.e., an absence of “constraint or compulsion having its basis in the natural order or constitution of things, esp. such constraint perceived as a law prevailing throughout the material universe and within the sphere of human action.”